\documentclass[
12pt,
3p,
review,
]{elsarticle}
\usepackage{hyperref}
\usepackage{graphicx,graphics}
\usepackage{placeins}
\usepackage{amsmath}
\usepackage{amssymb}
\usepackage{mathtools}
\usepackage{array}
\usepackage{hyperref}
\usepackage{color}
\usepackage{soul}
\usepackage{dcolumn}% Align table columns on decimal point
\usepackage{bm}% bold math
\usepackage{todonotes}
\usepackage[utf8]{inputenc}
\usepackage[english]{babel}
\usepackage{libertine}
\usepackage{pgfplots}
\usepgfplotslibrary{polar}
\pgfplotsset{compat=1.9}
\usetikzlibrary{calc}

\usepackage[title]{appendix}

\usepackage{multirow}

\usepackage{layouts}

\usepackage{caption}
\usepackage{subcaption}
\renewcommand{\l}{\mathopen{}\mathclose\bgroup\left}
\renewcommand{\r}{\aftergroup\egroup\right}

\let\originaleps=\epsilon
\let\epsilon=\varepsilon
\let\varepsilon=\originaleps
\usepackage{stmaryrd}

\mathchardef\hy="2D
\definecolor{mydark_blue}{RGB}{0, 0, 139}
\definecolor{myblue}{RGB}{0, 0, 255}
\definecolor{mycyan}{RGB}{0, 255, 255}  
\definecolor{mygreen}{RGB}{0, 255, 0}
\definecolor{myyellow}{RGB}{255, 255, 0}
\definecolor{myred}{RGB}{255, 0, 0}
\definecolor{mydark_red}{RGB}{139, 0, 0}
\definecolor{myblack}{RGB}{0, 0, 0}

\definecolor{BRY_1}{RGB}{  0,  0,255}
\definecolor{BRY_2}{RGB}{127,  0,127}
\definecolor{BRY_3}{RGB}{255,  0,  0}
\definecolor{BRY_4}{RGB}{255,127,  0}
\definecolor{BRY_5}{RGB}{255,255, 85}

\journal{XXXXXXXX}

\begin{document}

\begin{frontmatter}
\title{Interpretable MA-island clusters and fingerprints relating bainite microstructures to composition and processing temperature}

\author[mymainaddress]{Vinod Kumar\fnref{fn1}}
\author[mymainaddress]{Sharukh Hussain\fnref{fn1}}
\author[mysecondaryaddress]{Priyanka S}
\author[mymainaddress]{P G Kubendran Amos\corref{mycorrespondingauthor}}
\ead{prince@nitt.edu}

\cortext[mycorrespondingauthor]{P G Kubendran Amos}

\fntext[fn1]{The authors contributed equally.}

\address[mymainaddress]{Theoretical Metallurgy Group,
Department of Metallurgical and Materials Engineering,\\
National Institute of Technology Tiruchirappalli, \\
Tamil Nadu, India}

\address[mysecondaryaddress]{School of Electronics Engineering,
VIT-AP University,\\
Amaravati, Andra Pradesh, India
}

\begin{abstract} 

Realising the affect of composition and processing condition on bainite microstructures is often challenging, owing to the intricate distribution of the constituent phases.
In this work, scanning electron micrographs of non-isothermally transformed bainite, with martensite-austenite (MA) islands, are analysed to relate the microstructures to the composition and quench-stop temperature. 
The inadequacy of the MA-islands' geometric features, namely aspect ratio, polygon area and compactness, in establishing this relation is made evident from Kullback-Leibler (KL) divergence at the outset. 
Clustering the bainite microstructures, following a combination of feature extraction and dimensionality reduction, further fails to realise the affect of composition and processing temperature.
Deep-learning analysis of the individual MA islands, in contrast to the bainite microstructures, yields interpretable clusters with characteristically distinct size and morphology.
These five clusters, referred to as fine- and coarse-dendrite, fine- and coarse-polygon and elongated,  are exceptionally discernible and can be adopted to describe any MA island.
Characterising the bainite microstructures, based on the distribution of the interpretable MA-island clusters, generates \textit{fingerprints} that sufficiently relates the composition and processing conditions with the microstructures.  

\end{abstract}


\end{frontmatter}

% \linenumbers

\section{Introduction}

Of the different types of steels studied and employed in wide range of applications, bainite steels hold a unique place~\cite{krauss2015steels}. 
This is largely because, despite their extensive usage, microstructures of bainite steels are often complex and offer unclear understanding of the transformation mechanism~\cite{bhadeshia2001bainite}. 
The fact that it took years, and some debates, to gain a consensus in the definition of bainite is a testament to the intricacy of its microstructures~\cite{aaronson1990bainite,fielding2013bainite}. 
These intricacies are not restricted to the morphology of its features, which can be viewed as the product of a non-cooperative (competitive) growth of eutectoid phases, but also the existence of rather unlikely phases~\cite{bramfitt1990perspective}.

The processing (and/or heat treatment) technique along with composition contribute to the morphology and the nature of the phases constituting the bainite microstructures~\cite{lee1988mechanisms}. 
Bainite steel emerging from isothermal treatment can broadly be categorised as upper and lower, based on the distribution of carbides~\cite{yin2017morphology}.
Transformation at higher finish-cooling (or holding) temperature ensures adequate migration of carbon, thereby forming upper bainite characterised by the carbides at the boundaries of the ferrite structures. 
On the other hand, lack of adequate carbon mobility at low holding temperatures favours the formation of carbides within the ferrite structures, ultimately yielding lower bainite. 
This clear and straightforward distinction become ill-suited while describing bainite microstructures of non-isothermal treatment,~\cite{tian2019transformation}. 
Besides forming carbides both within and around ferrite, bainite microstructures resulting from continuous cooling accommodate additional secondary phases~\cite{huang2014secondary}.
The prevalence of the characteristic \textit{incomplete transformation}, during the non-isothermal treatment, is a primary cause for the introduction of additional phases. 
And the nature of these phases are essentially dictated by the alloying elements~\cite{wu2017incomplete,bhadeshia1982bainite}. 

Incomplete transformation in bainite steels, particularly resulting from non-isothermal treatment, is characterised by the significant deviation from expected phase-fraction (level rule).
Depending on the interpretation of bainite evolution, the incomplete transformation is attributed either to progressive decrease in driving force due to accumulation of carbon~\cite{caballero2013new} or solute drag~\cite{hillert1994diffusion}.  
Irrespective of the reason, by preserving the parent austenite, incomplete transformation facilitates in introducing a wide-range of secondary phases including pearlite~\cite{caballero2009new}, martensite~\cite{lu2021effect} and retained austenite~\cite{wang2012new}. 
Given that the introduction of secondary phases results in degenerate microstructures, in addition to conventional upper and lower bainite, the classification is extended to degenerate-upper and -lower bainite~\cite{zhang2016high}.

Phases introduced after the incomplete transformation are primarily dictated by the cooling regime, including final cooling temperature, and alloying elements. 
Correspondingly, the untransformed parent austenite evolves into one or more secondary phases, beside ferrite and carbide~\cite{zajac2005characterisation}. 
Slower cooling rate, or higher holding temperature, favours long-range migration of carbon.
Consequently, concentration gradient across the interface separating bainite-ferrite and austenite gets minimised, and formation of carbides is correspondingly regulated. 
Besides carbon, the thermal cycle associated with slow cooling rate (or higher holding temperature) enable the long-range diffusion of alloying elements. 
The enhanced migration of carbon and alloying elements rarely translate to a uniform distribution in austenite. 
Therefore, while the relatively carbon-rich regions in austenite get stabilised, the deficit pockets transform to martensite, ultimately introducing martensite-austenite (MA) island~\cite{zhang2002accurate}.
These MA island are a unique combination of secondary phases in bainite microstructures wherein the martensite is locally surrounded, and intertwined, by relatively carbon-rich austenite.
Despite it seemingly complex evolution, MA islands are not uncommon in bainite microstructures, particularly in those resulting from non-isothermal transformation. 
However, since the transformation depend on spatial distribution of carbon and other alloying elements, MA islands generally assume complex morphologies~\cite{park2001interpretation}.
Ultimately, the formation and the features of MA islands, including size and morphology are governed by composition, and thermal cycle dictating the migration of carbon and associated alloying elements. 
Considering that MA islands reflect the distribution of carbon and its alloying element in parent austenite,  \textit{a bainite steel of definite composition, exposed to a specific cooling regime, would often encompass varied shapes of MA islands.}
The significant influence of the MA islands on the mechanical properties of bainite steels demand a comprehensive analysis of their features including morphology.

The interface separating MA island from the surrounding matrix, owing to its characteristic decohesion, is viewed as a potential site for crack nucleation~\cite{avramovic2009effect}.
Correspondingly, in addition to yield strength~\cite{isasti2014microstructural} to bendability~\cite{kaijalainen2016influence}, MA islands affect the fracture toughness of bainite steel. 
The degree of this influence extend beyond the size of the MA islands to their morphology. 
Stated otherwise, the shape of the MA island by dictating the interface area alter the  the crack-initiation stress and the subsequent formation of void~\cite{taboada2019substructure}. 
Besides the formation, the propagation of a crack is also affected by its interaction with MA islands~\cite{mao2018relationship}. 
This interaction, which is primarily driven by the morphology and aspect ratio of the island, regulates the resistance offered by bainite steels to the growth of crack, and determines its propagating path. 
Given this influence, desired mechanical properties in bainite steels with MA islands can only be achieved by suitably varying their morphology and aspect ratio.
To that end, attempts are made in this work to analyse and relate the critical features of the MA islands with the composition and a processing parameter, called quench-stop temperature.

A conventional approach of relating morphology of a phase to processing condition or resulting properties involves considering different aspects of the shapes like  perimeter, aspect ratio, area and compaction. 
Correspondingly, the present work will begin with the attempts to relate the MA islands with composition and processing condition through the geometric parameters. 
Following the geometric investigation, the intricate morphology of the islands will be considered in its entirety. 
This comprehensive consideration of the MA island morphology is achieved by employing deep learning techniques. 
Recently, by labelling different bainite steels based on their Charpy Impact Value (CIV), features of the MA islands along with other components have \textit{collectively} been related to these alloys through deep-learning based classification technique~\cite{ackermann2023explainable}. 
In this work, on the other hand, the morphologies of the MA islands are individually analysed, to make them approached, and ultimately, relatable to the composition and processing conditions.

\begin{figure}
    \centering
      \begin{tabular}{@{}c@{}}
      \includegraphics[width=1.0\textwidth]{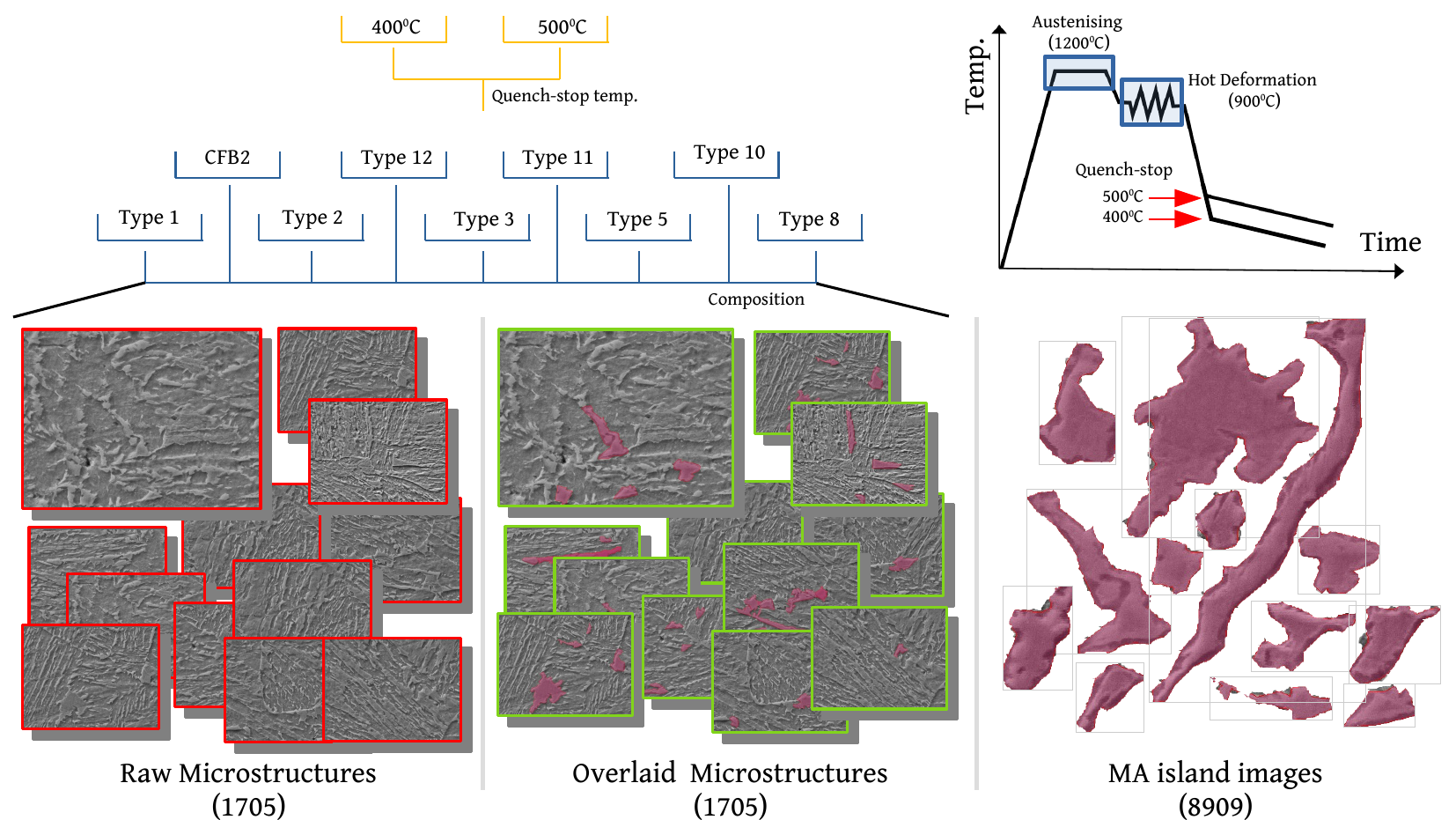}
    \end{tabular}
    \caption{ The heat treatment cycles imposed on the different types of steels with varying composition is schematically illustrated in top right corner. Three types of data, raw and overlaid microstructures along with isolated MA islands, are analysed in this work. These dataset encompass steels nine different composition processed under two different conditions.
    \label{fig:Dataset2}}
\end{figure}

\section{Principal dataset}

The dataset extensively analysed in the present work has separately been reported in Ref.~\cite{iren2021aachen}.

\subsection{Processing conditions}

Steel samples of eight different compositions in the form of forged ingot of dimension $20 \times \times 20 \times 65$ mm$^3$ were considered to generate this dataset. 
These ingots were austenised at $1200^o$C for 10 minutes before being hot rolled at $900^o$C by imposing a 0.3 applied strain and 10 s$-1$ applied strain. 
Following the deformation, a set of sample encompassing all compositions were cooled at the rate of $5$ K s$^{-1}$ to a quench-stop temperature of $400^o C$, while the rest were cooled to $500^o$ C at a slower rate of $2$ K s$^{-1}$. 
Upon reaching the quench-stop temperature, all the samples were cooled at $0.3$ K s$^{-1}$ to room temperature. 
A schematic representation of the thermal cycle is included in the top right panel of Fig.~\ref{fig:Dataset2}.
These heat-treatment cycles were devised to facilitate suitable bainite transformation. 
While the differences in the thermal cycle, including cooling rate, are referred to by their quenching-stop temperature in the subsequent discussions, the samples are distinguished as Type 1, 2, 3, 5, 8, 10, 11, 12, and CFB2 to indicate the disparity in the composition. 
After appropriate preparation, the micrographs of the samples were captured through Scanning Electron Microscopy (SEM). 
These micrographs of the heat-treated samples for a major portion of the principal dataset, as indicated in Fig.~\ref{fig:Dataset2}.

\subsection{Micrographs and MA island datasets}

\begin{figure}
    \centering
      \begin{tabular}{@{}c@{}}
      \includegraphics[width=0.75\textwidth]{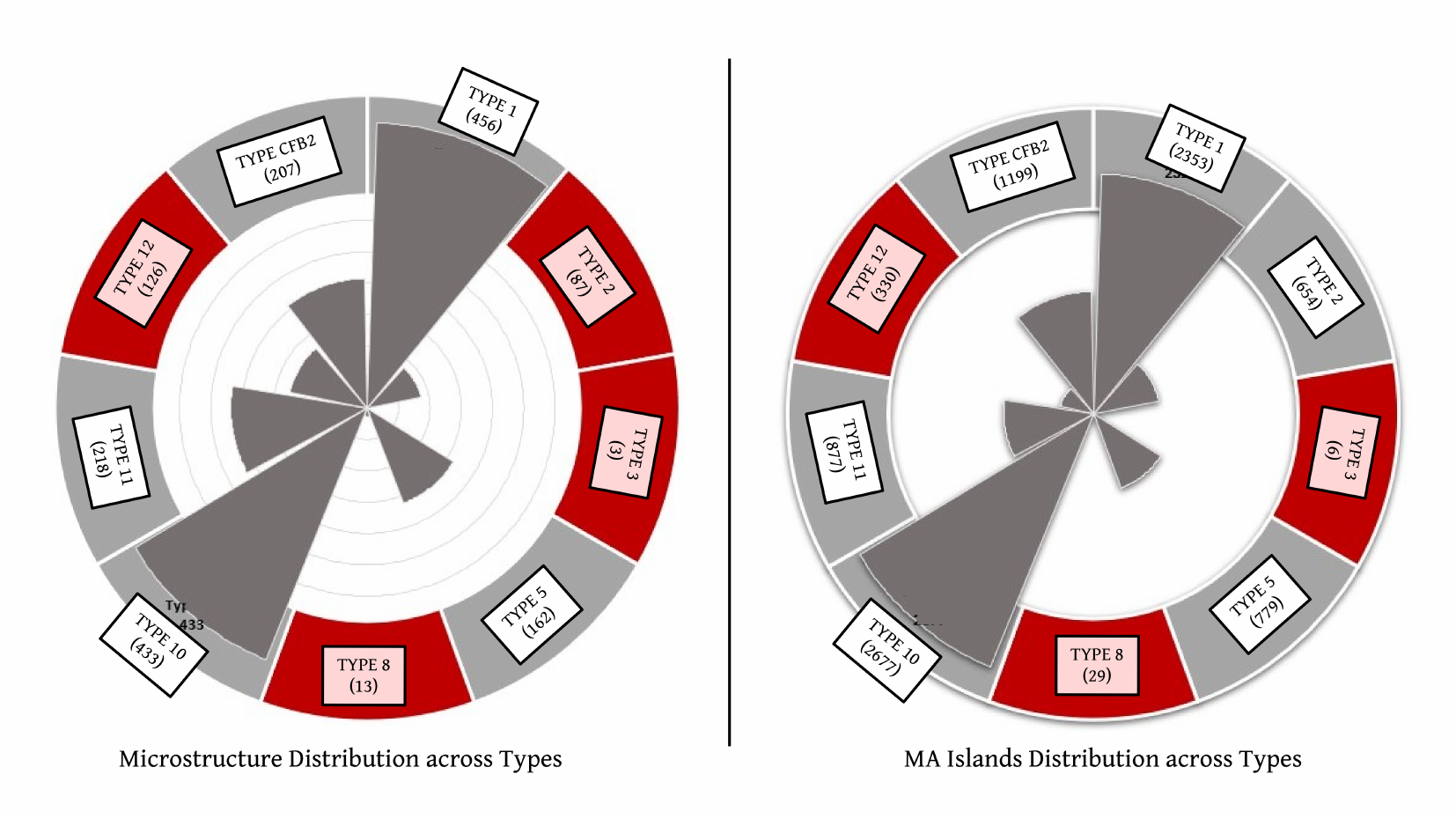}
    \end{tabular}
    \caption{ Radial histogram representation of the contribution made by individual types of steel to the microstructures and MA islands.
    \label{fig:work_dataset}}
\end{figure}

After appropriate preparation, the micrographs of the samples were captured through Scanning Electron Microscopy (SEM) at $4000 \times$ magnification. 
These micrographs of the heat-treated samples  amount to 1705 high-quality images, and form a major portion of the principal dataset, as indicated in Fig.~\ref{fig:Dataset2}.
Through human experts, 8909 points-of-interests with each indicating a MA island have been realised in these 1705 microstructures. 
The morphologies of the MA-islands are gleaned by the experts through the superimposition of polygons defined by suitable coordinates.
The points indicating the position of the MA islands and the coordinates capturing their morphologies are recorded through \textit{shapely}.
A duplicate of the \textit{raw} 1705 micrographs is generated by overlaying the point and polygon emerging from the coordinates. 
In the resulting set of micrographs, the MA islands are distinguished. 
This duplicate 1705 images with highlighted MA islands represented as a separate dataset in Fig.~\ref{fig:Dataset2}, and are referred to as overlaid microstructures. 
From the overlaid microstructures, the highlighted MA islands are isolated to form a third subset of data comprising exclusively of 8909  MA islands.
This dataset of MA islands, along with other sets of micrographs, are schematically represented in Fig.~\ref{fig:Dataset2}. 
Ultimately, the micrograph dataset studied in the present work includes raw and overlaid microstructures along with compilation of individual MA islands. 

Although unsupervised deep-learning techniques do not require training data, their performance still reflect the wealth of available information. 
Therefore, steels with significant number of micrographs are realised and separated out from present deep-learning analysis.
Types of steels which are overlooked due to their relatively minimal contribution to the micrograph dataset are distinguished in Fig.~\ref{fig:work_dataset}, wherein a graphical representation of steels based on the associated number of microstructures is offered. 
Correspondingly, micrographs pertaining to steels of type 2, 3, 8 and 12 are not involved in the subsequent analysis. 
Although type 2 steel offers less number of micrographs, the count of MA islands in these limited microstructures are reasonably higher. 
Accordingly, these MA islands associated with type 2 steels are included in the corresponding subset of data to enhance the wealth of information. 
The dataset excluding the types with limited representation accommodate \textit{1476 raw and overlaid microstructure} each with \textit{8539 MA islands}. 

\begin{figure}
    \centering
      \begin{tabular}{@{}c@{}}
      \includegraphics[width=1.0\textwidth]{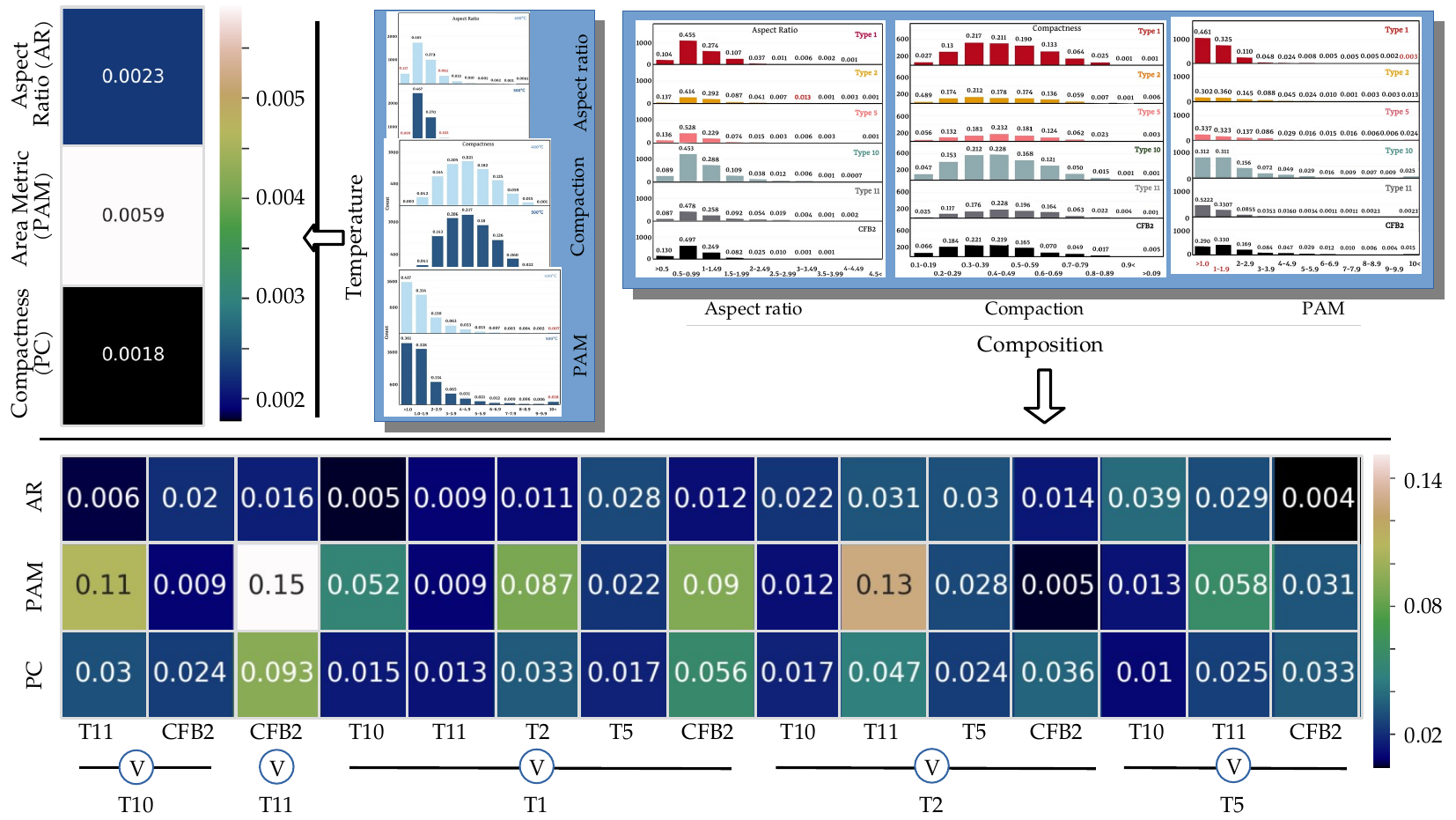}
    \end{tabular}
    \caption{ The number-density distributions of the MA island based on their geometric features in bainite microstructures of varying compositions and quench-stop temperatures are presented in top-right panel. Kullback-Leibler (KL) divergence quantifying the differences in the distribution of individual geometric-parameters across microstructures of varying processing condition is illustrated in top-left panel, while in the bottom KL divergence between the distribution of polygon compactness (PC), area metric (PAM) and aspect ratio (AR) is presented. 
    \label{fig:KLd}}
\end{figure}

\section{Conventional treatment}

In the existing studies processing parameters and composition have been related to the bainite microstructures through the geometric features of the MA islands~\cite{caballero2012influence}. 
Noticeable disparity in the area and perimeter of the MA islands, introduced by the differences in the cooling regime, is adopted to associate bainite microstructures with processing conditions~\cite{ackermann2020effect}. 
Therefore, before involving deep-learning techniques, geometric-parameters based conventional approach is extended to the current dataset to examine their efficacy. 

\subsection{Geometric data}

In addition to the micrographs, the overall dataset includes (meta)data holding numerical information on individual MA islands. 
These numerical data, while distinguishing individual micrographs through URL, include coordinate that describe the position and morphology of the MA islands. 
Based on the coordinates, the intricate shapes of the MA islands are reconstructed to determine the corresponding geometrical features.
The geometric parameters  considered in the present investigation include
\begin{enumerate}
 \item Aspect ratio: A dimensionless parameter describing the span of the MA island. It is ascertained by taking the ratio of the maximum length and width of the island,
 \item Polygon area metric (PAM): Area of the MA island in micrometers,
 \item Polygon compactness: A parameter cumulatively describing both area (A) and perimeter (P) of the MA islands through the relation $4\pi (A/P)$. 
\end{enumerate}
Given these parameters are reflective of the size and morphology of the islands, adopting them exclusively to analyse bainite microstructures seemingly translates to comprehending the effect of processing condition and composition on MA-islands.

The conventional treatment of relating bainite microstructures to the thermal cycle and composition begins with determining the range of values assumed by the different geometric features across the bainite microstructures. 
This numerical spectrum, which varies with the geometric parameters, is resolved into series of bins encompassing specific minimum and maximum values.
Based on their quantitative measurement, MA islands are placed in the corresponding bins of the geometric feature, ultimately yielding a number-density distribution represented in the form of the histogram. 
(Number density defined as the ratio of the number of MA islands associated with a bin and total number of islands in given categorised of bainite microstructures differentiated either based on quench-stop temperature or composition.)

\subsection{Inadequacy of geometric parameters}

In Fig.~\ref{fig:KLd}, the number-density distributions of MA islands, based on aspect ratio, polygon area matrics and compactness, across the bainite microstructures pertaining to a given quench-stop temperature or composition are illustrated separately.
Firstly, in complete agreement with the current understanding, the number-density distribution unravel that the quenching stopped at high temperature (500$^0$C) engenders larger MA islands when compared 400$^0$C, low temperature.
This disparity can be attributed to the enhanced carbon diffusion favoured by the characteristic low-rate from relatively higher temperature. 
Besides this difference in the size of the MA island, only few other minor deviations are observed in the distribution of number-density across microstructures of varying compositions and processing, irrespective of the geometric parameters. 
In other words, the outcomes of the present conventional treatment involving the geometric parameters, comprehensively discussed elsewhere, fails to offers sufficient characteristic differences to relate microstructures with the corresponding quench-stop temperature and composition. 

\begin{figure}
    \centering
      \begin{tabular}{@{}c@{}}
      \includegraphics[width=0.8\textwidth]{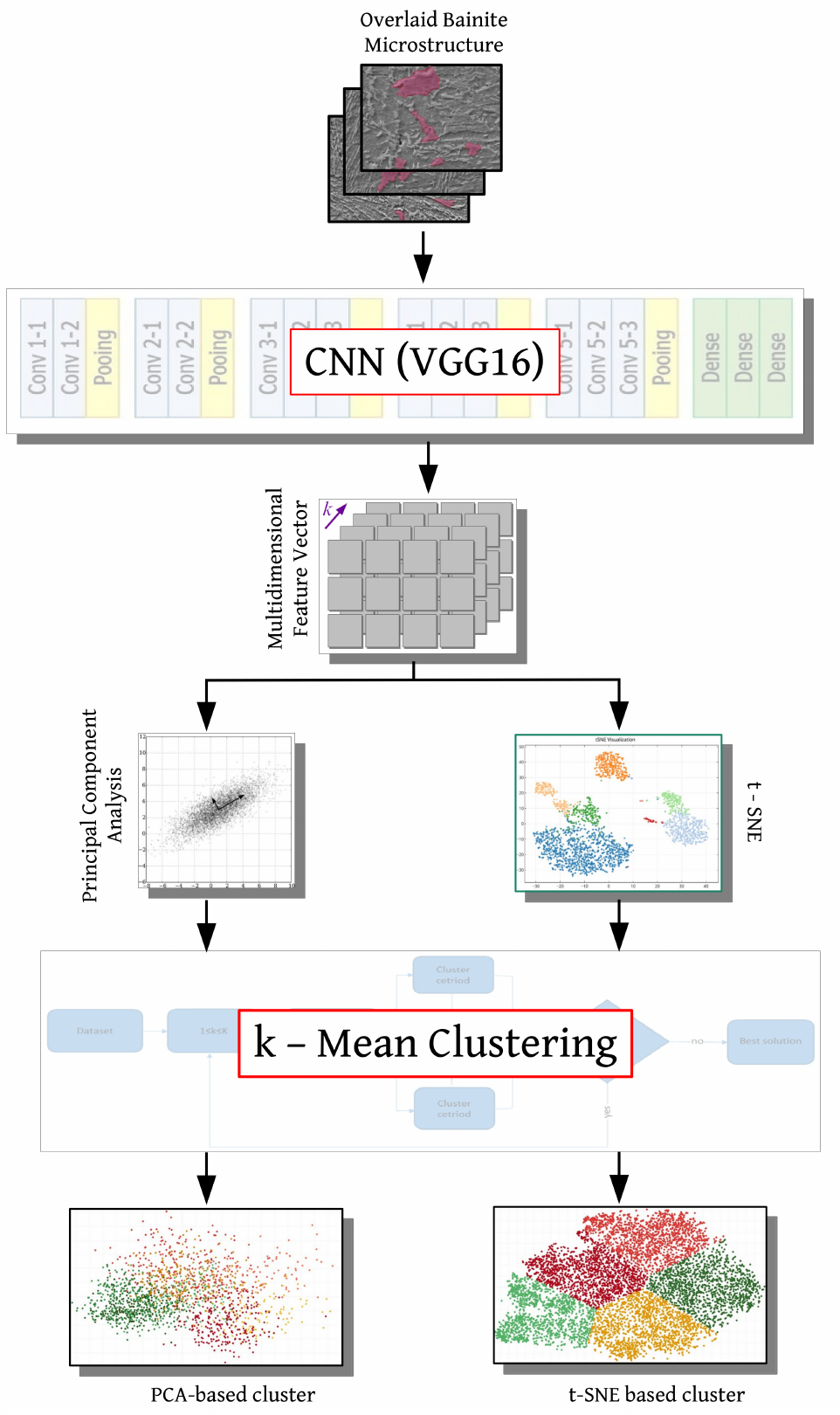}
    \end{tabular}
    \caption{ A deep-learning embedded framework adopted in the present investigation to relate the intricate bainite microstructures with transformation temperature and composition with the intent to focus on the morphology of the constituent phases.
    \label{fig:techniqe}}
\end{figure}

The inadequacy of the geometric-features based approach in associating bainite microstructures with the composition and processing condition is substantiated by quantifying the similarity in the resulting distributions.
This is achieved by estimating Kullback-Leibler (KL) divergence between all pairs of geometric-feature based distributions, in the top-right panel of Fig.~\ref{fig:KLd}, resulting from different categories of microstructures.
The mathematical formalism underpinning the KL divergence calculation is discussed in Appendix 1.
The numerical formulation of KL divergence, discussed in Fig.~\ref{fig:KLd}, indicate that this parameter quantifies the dissimilarity between two distributions.  
Put simply, larger the value of the KL divergence, greater is the disparity between distributions. 
The KL-divergence values of various sets of number-density distributions reflecting the size and morphology of the MA islands in bainite microstructures of specific composition or quench-temperature is illustrated in Fig.~\ref{fig:KLd}. 
Since there are only two quench-stop temperatures, 500$^0$C and 400$^0$C, KL divergence  is ascertained for the corresponding two distribution, thereby generating one distinct value for each geometric feature. 
On the top-left panel of Fig.~\ref{fig:KLd} these three KL divergence values are presented. 
In addition to two quench-stop temperatures, six different compositions are included in this investigation which are represented as T1,T2, T5, T10, T11 and CFB2. 
For estimating the corresponding KL divergence, fifteen different pairs of composition are realised for each geometric parameter, and their similarity is quantified.
In the lower half of Fig.~\ref{fig:KLd}, all the forty five KL divergence values calculated for three different geometric-parameter distribution across different composition is included. 
It is evident from Fig.~\ref{fig:KLd} that maximum KL divergence for composition and temperature distribution is $0.15$ and $0.0059$, respectively.
Given the significance and bounds of KL divergence, the values $0.15$ and $0.0059$ affirm the extensive similarity between the temperature and composition distributions of geometric features, thereby substantiating the inadequacy of the conventional treatment, and importantly, the need for relatively sophisticated deep-learning technique.

\section{Deep-learning analysis}

\subsection{Framework}

Given the inadequacy of geometric features and the complex shapes, a deep learning framework is developed and employed to wield the MA islands more effectively in relating the bainite microstructures to quench-stop temperature and composition. 
The deep-learning framework developed for a more detailed analysis of the MA islands and the associated bainite microstructures is illustrated in Fig.~\ref{fig:techniqe}.
Individual components of this framework are discussed in Appendix 2. 

Bainite microstructures with highlighted MA islands are converted to multilayer vectors through convolutional neural network (CNN). 
In a two-stage operation of CNN involving conversion and processing largely for classification, the latter is discarded to yield a more manageable numerical representation of the microstructure. 
Moreover, a neural network (VGG16) adequately pre-trained through transfer learning is involved to ensure an accurate grasp of the microstructural features.

Given the pixel size ($224 \times 224$) and the colour space of the bainite microstructure, the conversion using neural network yields a feature vector of 4096 dimension for every micrograph. 
The dimensions of these vectors, representing the key features of bainite microstructures, are significantly reduced to enhance the efficiency of the subsequent treatment.
Two different dimensionality-reduction techniques, offering respectively distinct pathways with minimal information loss, are adopted for the deep-learning analysis.
Principal component analysis (PCA) by projecting the feature-vector datapoints onto a subspace of fewer orthogonal axes reduces the dimension from 4096 to 451 for all bainite microstructures.
By involving probability-based measure of similarity over t-distribution, the other dimensionality-reduction treatment,  t-distributed Stochastic Neighbor Embedding (t-SNE), facilitates the representation of feature vectors in low dimensional space.

Vectors representing the MA islands overlaid microstructures, following the dimensionality reduction, are clustered. 
Though reduced, since the vectors remain multidimensional, the established k-mean technique is adopted to handle this exhaustive data.  
Despite the involvement of a single clustering technique, two distinct pathways, PCA+k-mean and t-SNE+k-mean, are introduced by the different dimensionality reduction treatments. 

\subsection{Resulting clusters}

\begin{figure}
    \centering
      \begin{tabular}{@{}c@{}}
      \includegraphics[width=1.0\textwidth]{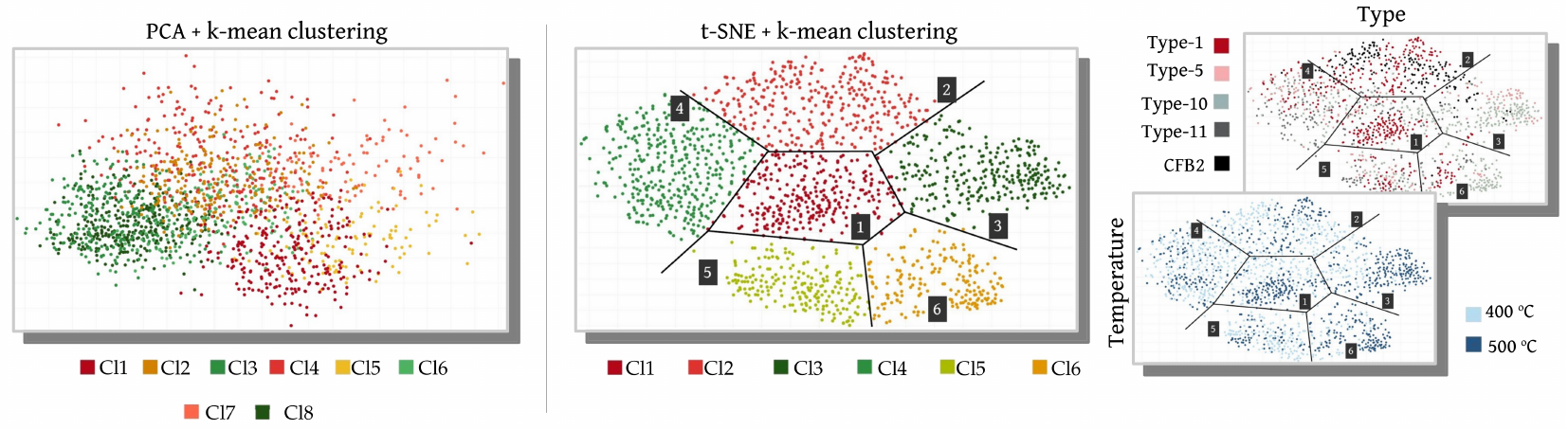}
    \end{tabular}
    \caption{ Clusters of feature vectors representing overlaid bainite microstructures generated by deep-learning framework. While in the left panel, the dimensions of the features vectors are reduced by Principal Component Analysis, t-SNE is adopted for representing the clusters in the middle window. The t-SNE + k-mean clusters are visualised based on the quench-stop temperature and composition in the right panel.
    \label{fig:PCA_tSNE}}
\end{figure}

The clusters of bainite microstructures generated by the k-mean algorithm, for different dimensionality reduction treatments, is shown in Fig.\ref{fig:PCA_tSNE}.
On the left panel, the clusters of microstructure feature-vectors with dimensions reduced through PCA is presented. 
The combination of PCA+k-mean delivers eight cluster which are distinguished by appropriate colour scheme in Fig.\ref{fig:PCA_tSNE}. 
These clusters, in contrast to the general expectation, are not distinct but significantly overlap with each other.  
Consequently, the clusters generated by PCA+k-mean fail to remain relevant in the subsequent analysis. 
Besides the algorithm, the lack clear demarcation between the clusters can be attributed to the prevalence of the MA islands with range of morphologies across the microstructures of different quenching conditions and compositions. 
The involvement of t-SNE, as illustrated in the middle panel of Fig.\ref{fig:PCA_tSNE}, spawns six visibly distinct clusters under the k-mean scheme. 
Since t-SNE essentially presents the multidimensional feature-vectors in low-dimensional space based on stochastic similarity, a better performance is offered over MA island highlighted microstructures with similarities, as indicated by KL divergence in Fig.~\ref{fig:KLd}.

Given the non-overlapping depiction of the feature-vector clusters under the combination of  t-SNE and k-mean, an approach is devised to relate the corresponding bainite microstructures with the quenching-stop temperature and composition.
Accordingly, the clusters the represented based on compositions (types) and quench-stop temperatures, through suitable colour scheme, and included in Fig.\ref{fig:PCA_tSNE}. 
A clear association of certain clusters to a specific quench-stop conditions or composition essentially unravels an underlying relation. 
However, Fig.\ref{fig:PCA_tSNE} indicates that all the six distinct clusters encompass microstructures of both quenching temperatures.

Though there are no exclusive association of clusters with the compositions,  three t-SNE + k-mean clusters predominantly accommodate microstructures of two types with traces of others.
Stated otherwise, Cluster 3, as labelled in Fig. 7b, predominantly includes microstructures from type-1 and CFB2, whereas type-5 and -10 dominate cluster 4. 
Similarly cluster 5 largely comprises of type-1 and -10 microstructures with noticeable traces of type-11.
Even though three clusters can largely be described based on the composition of the feature-vectors they constitute, there are two limitations that prevent the complete adoption of this approach to relate microstructures to the composition. 
Firstly, only three of the six clusters grant a type-based description, while the others remain a mixed bag of compositions. 
Secondly, and more importantly, even if there is an discernible association between all the clusters and the types (or temperatures) , the lack of any physical significance of these clusters hampers all attempts to meaningfully relate the corresponding microstructures to the processing conditions. 
Owing to these limitations, a the deep-learning approach is extended to generate \textit{microstructure fingerprints} for comprehending the affect of quench-stop temperature and composition on the overlaid bainite microstructures. 

\section{Microstructure fingerprint}

\begin{figure}
    \centering
      \begin{tabular}{@{}c@{}}
      \includegraphics[width=1.0\textwidth]{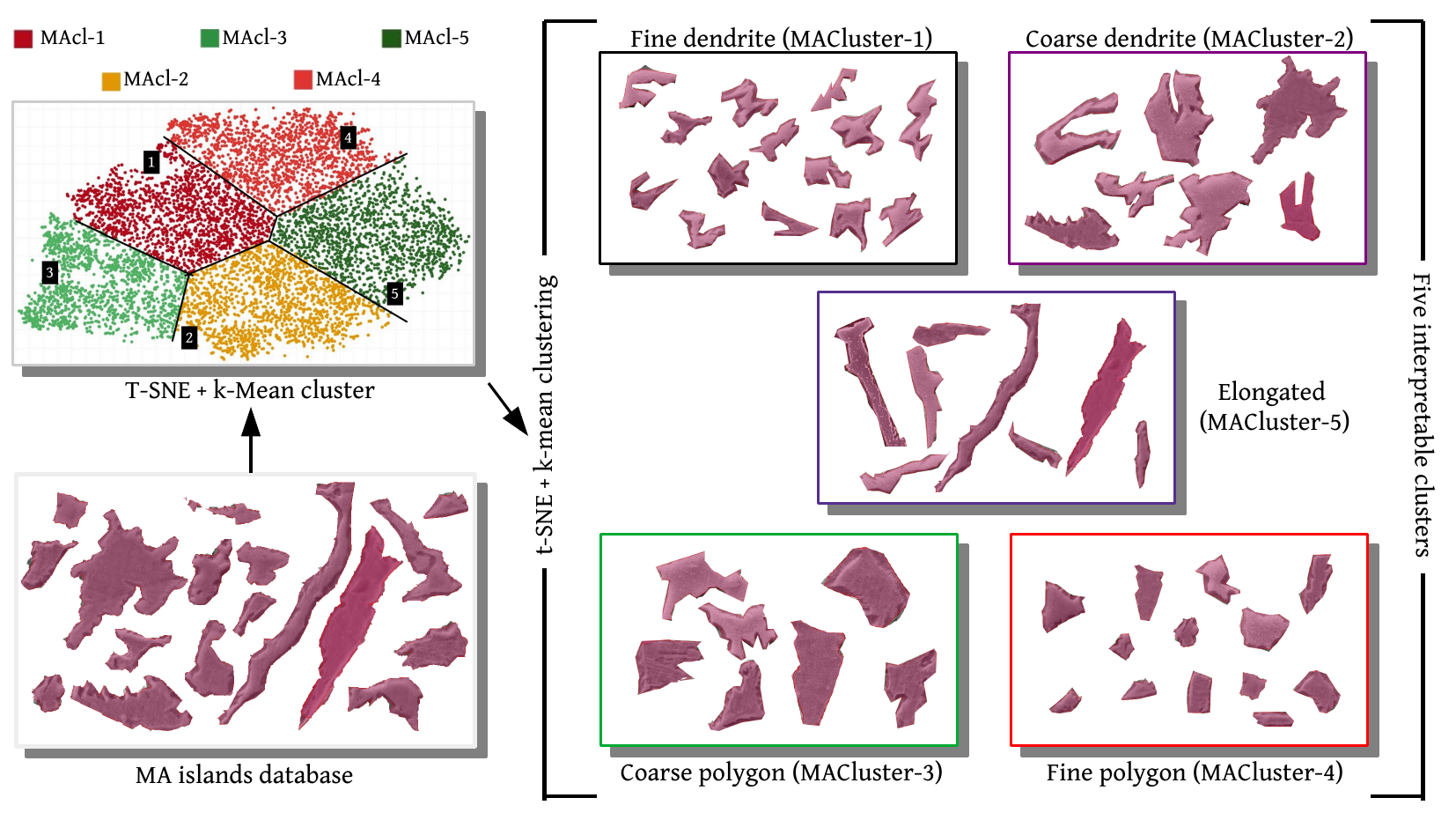}
    \end{tabular}
    \caption{ Clusters of MA islands generated by the deep-learning framework through the combination of t-SNE and k-mean clustering (Left). Visualisation of the MA islands that characteristically depict of each clusters (Right).
    \label{fig:MA_clusters}}
\end{figure}

\subsection{Clusters of MA islands}

The deep-learning framework involving the combination of t-SNE and k-mean cluster is extended analyse the dataset of individual MA islands, as opposed to the overlaid (or raw) microstructures. 
During this analysis, the MA islands are consciously handled to ensure that their characteristic features particular size are not in any way altered. 
In Fig.~\ref{fig:MA_clusters}, the deep-learning clusters emerging from combined treatment of t-SNE and k-mean clustering are included in the top left corner. 
Beneath the clusters, in Fig.~\ref{fig:MA_clusters}, a representation of the MA islands database is offered. 
In a straightforward attempt to interpret the usually non-interpretable deep-learning clusters, set of MA islands characteristically representing each clusters are realised and visualised. 
In other words, the MA islands of the feature vectors which are the centroid of each cluster, along with their neighbours, are traced and represented separately. 
These MA islands characteristically depicting each of the deep-learning clusters are shown in Fig.~\ref{fig:MA_clusters}. 
Based on the morphologies of the MA islands apparently constituting a cluster, a suitable suitable name is given to these clusters, in contrast to a generic reference. 
Correspondingly, deep-learning treatment of the MA islands database, through clustering, rather fortuitously distinguishes the islands as fine (MAcluster-1) and coarse dendrites (MAcluster-2), fine(MAcluster-4) and coarse polygons (MAcluster-3) along with elongated(MAcluster-5), as indicated in  Fig.~\ref{fig:MA_clusters}. 
Owing to the inadequacy of the microstructure classification, despite the involvement of reliable techniques, the seemingly interpretable clusters of MA islands are adopted to relate the bainite microstructures with the quench-stop temperature and composition.  

\subsection{Generating the fingerprint}

\begin{figure}
    \centering
      \begin{tabular}{@{}c@{}}
      \includegraphics[width=1.0\textwidth]{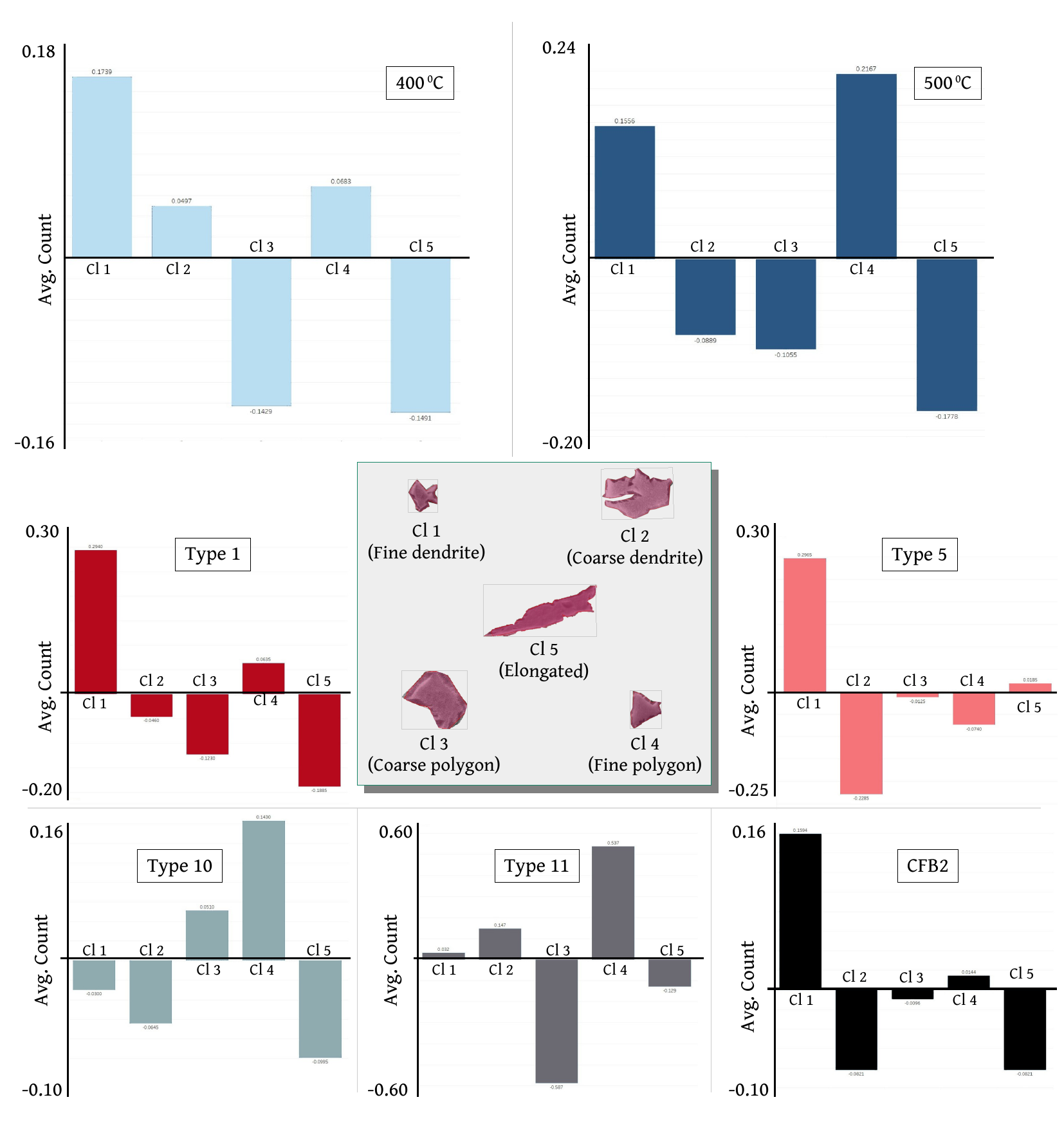}
    \end{tabular}
    \caption{ Fingerprints of the bainite microstructures resulting from (top) different quench-stop temperatures and (bottom) compositions (steel types), generated based distribution of the interpretable clusters included in the center.
    \label{fig:Finger}}
\end{figure}

The clusters, realised in Fig.~\ref{fig:MA_clusters}, are treated as \lq bag of features\rq \thinlines and all bainite microstructures in the raw dataset are described based on these morphologically distinguished MA islands.
The distribution of the MA islands across the bainite microstructures quenched at different temperatures and composition is represented based on the interpretable clusters in Fig.~\ref{fig:Finger}. 
In this illustration, the counts are indicated with respect to the average number of MA islands in a given class of bainite microstructures, such that the positive and negative values correspond to the above and below average population.
This representation of the bainite microstructures through the normalised count of morphologically-clustered MA islands is referred to as the \textit{fingerprint}.

The microstructure fingerprints of the bainite steels quench-stopped at two different temperatures, in Fig.~\ref{fig:Finger}, unravel that lower temperature introduces greater number (or above average) number of cluster-2 MA-islands when compared to processing at higher temperature.
Stated otherwise, by estimating the count of the coarse dendrite MA island, the bainite microstructures can be related to the quench-stop temperature. 
Similar to the quench-stop temperature, there are characteristic features in the fingerprints of the bainite microstructures with varying composition. 
These features in the fingerprints are adopted to relate the microstructure to the composition. 

It is evident from Fig.~\ref{fig:Finger}, type 10 is the only bainite steel with visibly above-average count of coarse-polygon MA island (cluster 3). 
Consequently, microstructures of type 10 can be characteristically described by the relative greater amount of coarse-polygon islands.
Microstructures of type-5 bainite, on the other hand, accommodate almost average number of  both coarse-polygon and elongated MA islands. 
Moreover, Fig.~\ref{fig:Finger}~ unravels that microstructures of type-11 steel are densely populated with MA islands of fine polygon morphology, as opposed any other types.
Though type-1 and CFB2 seem to yield similar fingerprint, MA islands with coarse dendrite morphology is extremely rare in the latter. 
Correspondingly, the distinction between these composition can be made by the relative presence of coarse-dendrite MA islands. 
Ultimately, microstructural fingerprints, when compared to direct clustering and extensive analysis geometric parameters, offer a convincing approach for relating microstructures to the processing conditions (quench-stop temperatures) and composition. 

\section{Conclusion}

The complex mechanism of bainite transformation results in the introduction of numerous phases. 
These phases generally assume intricate morphology ultimately posing a genuine challenge in relating the bainite microstructures to the corresponding processing conditions and/or composition. 
Even when the microstructures include a distinct phases like the MA islands, comprehending the affect of composition or temperature from the variations introduced in the geometric parameters including aspect ratio, polygon area and compactness noticeably fall short. 
Moreover, involvement of deep learning techniques through clustering and appropriate visualisation, while retaining the focus on the MA islands, fails to establish a convincing relation between the microstructures and composition (or process). 
Though it is evident that chemical composition and processing conditions influence the size and morphology of the phases, geometric factors and holistic consideration of the microstructures are inadequate to unravel the characteristic variations. 
Realising and clustering the MA islands through the deep-learning technique, on the other hand, facilitate in generating microstructure fingerprints, which by collectively involving size and morphology of the distinct phase, visibly distinguish microstructures based on the composition and processing temperature. 
Moreover, the clustering of the MA islands yields a comprehensible classes which can be extended to similar investigations. 

\section*{Appendices}

\subsection*{Appendix 1: Kullback-Leibler (KL) divergence}

In this work, Kullback-Leibler (KL) divergence is adopted to quantify the similarity (or differences) between curves indicating the distribution of MA islands with specific geometric features across bainite microstructures of varying composition (or processing condition). 
This is achieved by measuring the deviation of a given distribution in view of reference. 
Therefore, KL divergence essentially quantifies the similarity between pairs of distributions. 

For For two discrete probability distributions P and Q over a shared probability space X, KL divergence quantifies the expected extra bits required to encode data drawn from $P$ when using a code optimized for $Q$. 
This divergence is mathematically expressed as
\begin{equation}\label{eq:av}
 D_\text{KL}(P||Q) = \sum_{x\in X}P(x)\text{log}\frac{P(x)}{Q(x)},
\end{equation}
where $P(x)$ and $Q(x)$ represent the probability of an event $x$ occurring in distribution $P$ and $Q$, respectively. While the ratio $P(x)$ and $Q(x)$ expresses the likelihood of  $x$ in $P$ when compared to $Q$, the logarithmic estimation, weighted by $P(x)$, and its summation of all possible $x$, gives the amount of information (in bits) required to distinguish $P$ from $Q$.

\subsection*{Appendix 2: Components of deep-learning framework}

Given the intricate shape of MA islands, it is rather challenging to accurately determine these factors. 
Stated otherwise, even the investigations based on conventional parameters can be made more approachable when human expertise is augmented with advanced deep-learning or object-detection technique~\cite{prabakar2023regression,venkatanarayanan2023accessing}. 
Taking cognisance of its potential, attempts are made using deep learning to decipher and relate complex bainite microstructure, and MA islands images, with composition and processing temperature with minimal human efforts. 

One of the well-known deep learning technique, necessitating less human intervention \textit{id est} unsupervised, is clustering. 
When exposed to unlabeled dataset of images, this approach realises latent pattern in the features, and clusters similar datapoints together.
Bainite microstructures, with highlighted MA islands, are initially examined through the deep learning techniques, in the present work, to cluster them based on their likeness. 
Similar approach is subsequently adopted to examine the sectioned-out individual MA islands exclusively, while overlooking the other phases of the microstructure.  
Clustering is adopted encouraged by the existing reports, that demonstrate association specific groups to a set of processing parameters~(\cite{holm2020overview} and references therein).
Given clustering is based on the morphology of the phases in the microstructures, accurate extraction of the associated features plays a critical role. 
In order to accurately extract the characteristic features of the microstructures and MA island images, convolutional neural networks (CNNs) are employed.  

\subsubsection*{Feature-vector extraction}

In conventional image analysis, the approach adopted by convolutional neural networks broadly consists of two steps - conversion and processing.
The first step involves translating a given image into multilayer vector encompassing its characteristic features. 
Analysing and treating the image solely based on its numerical representation is the subsequent step.
Classification of images using CNNs is a prime example of this approach. 
Encoding of images to vectors begins with a systematic pixel-by-pixel scanning by CNNs using a set of filter patches. 
During this raster, the imposed filters get activated based on the features of the image.  
The activation of the filter across the image is definitively recorded. 
This record of the filter activation render by the initial raster forms the first convolution layer. 
Scanning with different sets of filters hierarchically follows, ultimately generating progressive convolutional layers.
The varied activations resulting from several convolutions is \textit{rectified} and \textit{pooled} together to translate pixels into a vector. 
Generation of the vectors reflecting the features of the images signals the end of the first step. 
Considering the characteristic features of the images are captured in the vectors through the activation of the appropriate filters, this stage essentially contributes to feature extraction. 
Based on the vectors representing the characteristic features, microstructures are clustered in the present work.
Stated otherwise, the second step of processing the images is completely overlooked, by exclusively adopting CNNs to generate feature vectors. 

The excessive demand for data to design and train CNN is addressed by \textit{transfer learning}.  
Owing to the considerable overlap in features like textures, edges and blobs across different images including complex microstructures, a pre-trained network called VGG16 is adopted for the current studies. 
Trained over millions of images on the ImageNet dataset, this VGG16 network is relied to render accurate feature vectors of bainite microstructure and MA island images. 
% The architechure of the VGG16 adopted in this works shown in \cit. 
The architecture of the VGG16 network includes thirteen convolution layers, and three dense layers along five max-pooling layers. 
Of the layers constituting VGG16 network, characteristic weights are associated with sixteen. 
Given the primary focus of the convolution network, softmax layers related to processing of the feature vectors are removed. 

\subsubsection*{Dimensionality reduction}

Bainite microstructures and MA island images are resized to 224 $\times$ 224 pixels, and mapped onto the RGB colour space for feature extraction through VGG16 network . 
The underlying architecture of the adopted approach translates each microstructure to a feature vector of 4096 dimension.   
Despite the significant advancements in visualisation techniques, representing such extremely high-dimensional vectors and interpreting the resulting depiction is an improbable task.
Consequently, the dimensionality of the feature vectors representing the microstructures and MA islands is reduced by employing suitable approaches. 
Considering the intricate morphology of the phases in bainite microstructure, two different dimensionality-reduction treatments are separately involvement before the datapoints are clustered. 

\paragraph*{Principal Component Analysis}

Of the various exploratory tools capable of reducing the dimensions of a dataset, principal component analysis (PCA) is a well-established and widely used technique. 
This approach, while scaling down the dimensionality of the data, additionally attempts to minimise the resulting loss of information. 
The reduction of dimension, in the framework of PCA, can be viewed as the projection of datapoints on a subspace represented by fewer number of orthogonal axes, when compared to the original vector space.
The characteristic orthogonal axes describing the subspace of reduced dimension are called the principal components. 

An extensive focus on the variability of the datapoints, \textit{id est} microstructure feature-vectors, ensures the marginal loss of information, while identifying the principal components.
Stated otherwise, the approach of realising principal components centers around finding the eigen-components of the covariant matrix of the original dataset.
The emerging eigenvalues and eigenvectors, in combination with suitable Lagrange multiplier, fundamentally contribute to realising the variables with minimised dimensions, which can subsequently be employed to handle the original data with minimal loss of information. 

\paragraph*{t-distributed Stochastic Neighbor Embedding}

An alternate technique adopted in the present work for representing the multidimensional feature-vectors of the microstructure is the t-distributed Stochastic Neighbor Embedding (t-SNE) technique. 
While PCA relies on the variability of the datapoints to render a seemingly accurate representation in the low-dimensional space, local similarity between the data is central focus of t-SNE. 
In other words, as opposed to linear transformations involved in PCA, t-SNE is based on non-linear operations that preserves the local structure and similarity of the datapoints.
Moreover, unlike PCA, t-SNE is a stochastic approach wherein the representation of multidimensional dataset is based on the probability-based measure of similarity.

Without altering the characteristic space of the microstructure feature-vectors, in t-SNE approach, the multidimensional Euclidean distance between the datapoints is ascertained.
The estimated high-dimensional Euclidean distances are subsequently adopted to calculate the pairwise probability of each feature vector in the neighbourhood of other across the entire dataset.
The proximity of the vectors indicated by the conditional probability translates to their similarity.
While a definite parameter is used define the neighbourhood, the probabilistic calculations in the original space are based on Gaussian distribution. 
This approach is extended, by replacing Gaussian distribution with t-distribution, to map the microstructure feature-vector in a low-dimensional space.
Positions of the datapoints (feature vectors) in the reduced-dimensional space reflects the stochastic similarity characteristically estimated in t-SNE. 
Any mismatch in the mapping to the low dimensional space is minimised by progressively reducing  the sum of Kullback-Leibler divergences through gradient descent.

\subsubsection*{k-means clustering}

Feature vectors of bainite microstructures (and MA island images) with its dimensionality reduced by PCA or t-SNE is clustered based on their similarity. 
Though similarity between the datapoints is implicitly considered in t-SNE approach, it is vital to note that the underlying framework largely overlooks the dissimilarities across the different subgroup. 
Therefore, clustering is additional imposed on feature vectors represented by t-SNE.

One of the well-known clustering techniques, adopted for handling multidimensional dataset is the k-means clustering. 
The underlying formulation presents this clustering scheme as a variant of Gaussian (normal) mixture models.
A set a data can viewed as a combination of several distributions with different mean and variance. 
In mixture models, subgroups indicating individual distributions in the population is realised.
While the association of a datapoint with a cluster is probabilistic in mixture-models, it is absolutely definitive in k-means clustering. 
In other words, a data will exclusively belong to a cluster and is not shared between the subgroups in k-means clustering. 
This rigorous placement of datapoints in k-means clustering, is based on multidimensional calculation of Euclidean distances which render similar datapoints closer to one another. 
The subgroups are made distinct by the attempts to minimise the variance from the centroid. 
In the present work, this k-means algorithm is employed to cluster the feature-vectors of bainite microstructures and MA islands.

\section*{Declaration of Interest}
The authors declare that they have no known competing financial interests or personal relationships that could have appeared to influence the work reported in this paper.

\section*{Acknowledgments}

PGK Amos thanks the financial support of the SCIENCE \& ENGINEERING RESEARCH BOARD (SERB) under the project SRG/2021/000092.

\section*{Data availability}
The raw data required to reproduce these findings are available to download from \url{https://doi.org/10.6084/m9.figshare.c.5185004}

% \section*{References}

% \bibliographystyle{elsarticle-num}
% 
% \bibliography{library.bib}

\end{document}